\pdfoutput=1
\documentclass[11pt]{article}

\usepackage[utf8]{inputenc}
\usepackage[T1]{fontenc}
\usepackage{lmodern}
\usepackage[margin=1in]{geometry}
\usepackage{microtype}

\usepackage{amsmath,amssymb}
\usepackage{graphicx}
\usepackage{xcolor}
\usepackage{caption}
\usepackage{enumitem}
\usepackage{fancyhdr}
\usepackage{hyperref}
\usepackage{listings}

\usepackage{tikz}
\usetikzlibrary{arrows.meta,positioning,shapes.geometric,fit,backgrounds,calc}

\definecolor{convred}{RGB}{220,0,0}
\definecolor{headingblue}{RGB}{47,84,150}

\hypersetup{
  colorlinks=true,
  linkcolor=headingblue,
  urlcolor=headingblue,
  citecolor=headingblue,
  pdftitle={On Thread Convergence},
  pdfauthor={Vinod Grover and Manjunath Kudlur}
}

\usepackage{sectsty}
\allsectionsfont{\color{headingblue}}

\lstdefinestyle{algo}{
  basicstyle=\ttfamily\small,
  keywordstyle=\bfseries,
  commentstyle=\itshape,
  showstringspaces=false,
  columns=fullflexible,
  keepspaces=true,
  breaklines=true,
  frame=none,
  language=C,
  morekeywords={each,in,Unknown,Convergent}
}

\tikzset{
  fg/.style   = {draw, ellipse, minimum width=9mm, minimum height=6mm,
                 inner sep=1.5pt, font=\small},
  fgred/.style = {fg, draw=convred, text=black, very thick},
  cedge/.style = {->, >=Stealth, convred, very thick},
  uedge/.style = {->, >=Stealth},
  clabel/.style = {font=\scriptsize, inner sep=1pt}
}

\newcommand{\cmark}{\ensuremath{c}}

\title{\bfseries On Thread Convergence%
  \thanks{This note was written in April 2009 and was not published at the
  time. It is posted here in 2026, essentially unchanged, as a historical
  record of an early approach to thread-convergence analysis for GPUs; its
  terminology and hardware references (e.g.\ Tesla, \texttt{SSY}/\texttt{NOP.S})
  should be read in that 2009 context.}}
\author{%
  Vinod Grover\thanks{NVIDIA Corporation. Corresponding author:
    \href{mailto:vgrover@nvidia.com}{\texttt{vgrover@nvidia.com}}.}
  \qquad
  Manjunath Kudlur\thanks{This work was done while the author was at NVIDIA
    Corporation; the author is now at OpenAI
    (\href{mailto:keveman@openai.com}{\texttt{keveman@openai.com}}).}}
\date{April 2009}

\begin{document}

\maketitle

\begin{abstract}
We introduce a notion of \emph{convergence} for the nodes and edges of a
control-flow graph that captures whether a barrier placed at that location is
guaranteed to synchronize all threads of a thread block in every execution.
Convergence analysis lets a compiler determine when a barrier lies in a
uniformly executed region and therefore avoid the code transformations
otherwise required to implement thread-block barriers correctly on
warp-synchronous hardware. We formalize convergent nodes, convergent edges, and
well-synchronized programs; give two inference rules---a branch rule and a
merge rule; and present a linear-time iterative work-list algorithm that
propagates convergence information bidirectionally through the flow graph. We
then describe refinements that improve precision using single-entry
single-exit region information, path information, and thread-variance
information.
\end{abstract}

\tableofcontents
\bigskip

\section{Introduction}

In this document we present and motivate a notion of convergence for
flow-graph edges and nodes. We also present an algorithm for statically
computing whether a node or an edge is convergent for well synchronized
programs.

\subsection{Motivation}

Consider the following simple flow graph resulting from the use of a barrier
that is controlled by a short circuit conditional.

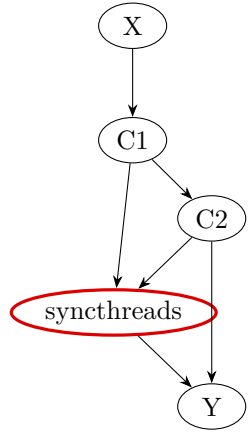
\begin{figure}[h]
  \centering
  \begin{tikzpicture}[node distance=9mm and 6mm]
    \node[fg] (X)  {X};
    \node[fg] (C1) [below=of X] {C1};
    \node[fg] (C2) [below right=6mm and 4mm of C1] {C2};
    \node[fgred] (S) [below left=8mm and 0mm of C2] {syncthreads};
    \node[fg] (Y)  [below right=8mm and 0mm of S] {Y};

    \draw[uedge] (X)  -- (C1);
    \draw[uedge] (C1) -- (C2);
    \draw[uedge] (C1) -- (S);
    \draw[uedge] (C2) -- (S);
    \draw[uedge] (C2) -- (Y);
    \draw[uedge] (S)  -- (Y);
  \end{tikzpicture}
  \caption*{}
\end{figure}

This executes incorrectly on our current GPUs since the hardware
implementation of the barrier works at warp level and not at thread level.
This can happen if some threads in the warp take the edge
$C1 \rightarrow \texttt{syncthreads}$ and the remaining threads of the warp
take the edge $C2 \rightarrow \texttt{syncthreads}$. If the compiler generates
a \texttt{SSY} instruction before C1 and a corresponding \texttt{NOP.S} at the
start of Y then a race results, since the \texttt{syncthreads} is executed at
the warp level.

In order to generate correct code for this for Tesla the compiler must
generate the following code.

\begin{figure}[h]
  \centering
  \begin{tikzpicture}[node distance=8mm and 6mm]
    \node[fg] (F0) {flag=0};
    \node[fg] (SSY) [below=of F0] {SSY(Y)};
    \node[fg] (C1)  [below=of SSY] {C1};
    \node[fg] (C2)  [below right=6mm and 5mm of C1] {C2};
    \node[fg] (F1)  [below left=7mm and 0mm of C2] {flag=1};
    \node[fg] (L)   [below right=7mm and 0mm of F1] {L: NOP.S};
    \node[fgred] (S) [below left=8mm and 0mm of L] {syncthreads};
    \node[fg] (Y)   [below right=8mm and 0mm of S] {Y};

    \draw[uedge] (F0)  -- (SSY);
    \draw[uedge] (SSY) -- (C1);
    \draw[uedge] (C1)  -- (C2);
    \draw[uedge] (C1)  -- (F1);
    \draw[uedge] (C2)  -- (F1);
    \draw[uedge] (F1)  -- (L);
    \draw[uedge] (C2)  -- (L);
    \draw[uedge] (L)   -- node[clabel,left] {1} (S);
    \draw[uedge] (L)   -- node[clabel,right] {0} (Y);
    \draw[uedge] (S)   -- (Y);
  \end{tikzpicture}
  \caption*{}
\end{figure}

In general, the presence of barriers in divergent single entry single exit
regions requires such a transformation. However if it was known that the edges
entering the barrier node were taken by all the threads of the thread-block
then the transformation can be avoided. In this paper we develop a method for
determining when edges are executed by all threads, also known as convergence
analysis.

\subsection{Definitions}

First we present some definitions that formalize the notions of convergence
and well synchronized programs.

\paragraph{Definition.} A flow-graph node $x$ is defined to be
\underline{\emph{convergent}} iff a barrier placed at $x$ will never fail to
synchronize all threads in the thread block in any execution.

\paragraph{Definition.} A flow-graph edge $e$ is defined to be
\underline{convergent} iff a barrier placed at $e$ (by splitting the edge)
will never fail to synchronize all the threads in the thread block, in any
execution.

\paragraph{Definition.} A flow-graph $G$ is said to be
\underline{\emph{well synchronized}} iff all barrier statements in $G$ will
never fail to synchronize all threads in any execution.

\medskip
Without any loss of generality, we assume that a barrier statement occurs in
its own basic block node for a given flow-graph.

The problem we consider is to statically determine which edges and nodes of a
flow graph are convergent. We describe an algorithm which will return the state
unknown (U) or convergent (C) for any given node or edge of the flow-graph.
This is represented as a vector called \emph{state} which maps a node or edge
to its convergence state.

\section{Convergence Inference}

The basic ideas behind the algorithm are contained in the following two simple
rules.

\begin{itemize}[leftmargin=2em]
  \item \textbf{Branch Rule:} A node $x$ is convergent iff all edges emanating
        from $x$ are convergent.
  \item \textbf{Merge Rule:} A node $x$ is convergent iff all edges incident on
        $x$ are convergent.
\end{itemize}

These rules can be used to infer the convergent states of a node or an edge
involved in a merge or a branch.

\subsection{Branch Inference}

Let's first consider a branch node, say $B$, and the set of edges, say $E$,
emanating from $B$. If among the states for $B$ and the state of each of the
edges in $E$ exactly one state is unknown and the rest are known to be
convergent then the unknown state can be assumed to be convergent as well.

We illustrate the branch rule inference in the following figure.

\begin{figure}[h]
  \centering
  \begin{tikzpicture}[baseline]
    \begin{scope}
      \node[fg] (b) {};
      \draw[cedge] (b) -- ++(-1,-1) node[clabel,pos=0.6,left] {\cmark};
      \draw[cedge] (b) -- ++(1,-1)  node[clabel,pos=0.6,right] {\cmark};
    \end{scope}
    \draw[-{Stealth[length=4mm]},very thick] (2,-0.5) -- (3.4,-0.5);
    \begin{scope}[xshift=5cm]
      \node[fg] (b2) {\cmark};
      \draw[cedge] (b2) -- ++(-1,-1) node[clabel,pos=0.6,left] {\cmark};
      \draw[cedge] (b2) -- ++(1,-1)  node[clabel,pos=0.6,right] {\cmark};
    \end{scope}
  \end{tikzpicture}
  \caption{Inferring convergence of a node using the branch rule}
\end{figure}
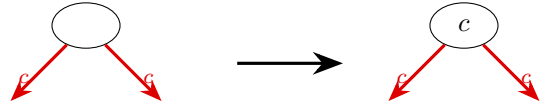

On the left hand side we have a branch node, with an unknown state, which has
2 successor edges known to be convergent. We can infer the state of the branch
node to be convergent as shown on the right. Similarly in the following figure,
we show how the branch rule can be used to infer the state of an edge.

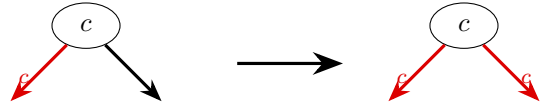
\begin{figure}[h]
  \centering
  \begin{tikzpicture}[baseline]
    \begin{scope}
      \node[fg] (b) {\cmark};
      \draw[cedge] (b) -- ++(-1,-1) node[clabel,pos=0.6,left] {\cmark};
      \draw[uedge,very thick] (b) -- ++(1,-1);
    \end{scope}
    \draw[-{Stealth[length=4mm]},very thick] (2,-0.5) -- (3.4,-0.5);
    \begin{scope}[xshift=5cm]
      \node[fg] (b2) {\cmark};
      \draw[cedge] (b2) -- ++(-1,-1) node[clabel,pos=0.6,left] {\cmark};
      \draw[cedge] (b2) -- ++(1,-1)  node[clabel,pos=0.6,right] {\cmark};
    \end{scope}
  \end{tikzpicture}
  \caption{Inferring convergence of an edge using the branch rule}
\end{figure}

On the left side of the figure, a branch node and only one of its 2 successor
edges is known to be convergent. We can infer the only unknown edge to be
convergent as well.

We also note that the branch rule can be used to infer convergence when the
branch node has a single successor. The following figures illustrate the branch
rule for such situations.

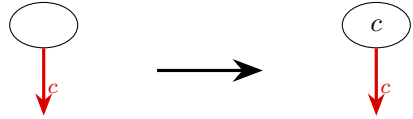
\begin{figure}[h]
  \centering
  \begin{tikzpicture}[baseline]
    \begin{scope}
      \node[fg] (b) {};
      \draw[cedge] (b) -- ++(0,-1.2) node[clabel,pos=0.6,right] {\cmark};
    \end{scope}
    \draw[-{Stealth[length=4mm]},very thick] (1.5,-0.6) -- (2.9,-0.6);
    \begin{scope}[xshift=4.4cm]
      \node[fg] (b2) {\cmark};
      \draw[cedge] (b2) -- ++(0,-1.2) node[clabel,pos=0.6,right] {\cmark};
    \end{scope}
  \end{tikzpicture}
  \caption{Inferring convergence of a unconditional branch node using the branch rule}
\end{figure}

We further note that branch rule inference propagates information in forward or
backward direction; either from a set of successor edges to the common
predecessor node or from a node to one of the successor edges.

\subsection{Merge Inference}

We now show how the merge rule can be used. Say we have a merge node $M$ and the
set of edges $E$ incident on it. If among the states of $M$ and each of the
states of edges in $E$ exactly one state is unknown and the rest are convergent
the unknown state can be assumed to be convergent as well. We illustrate the
merge rule in the following 2 figures.

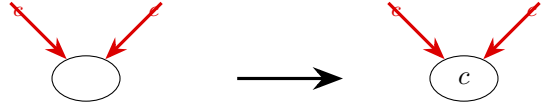
\begin{figure}[h]
  \centering
  \begin{tikzpicture}[baseline]
    \begin{scope}
      \node[fg] (m) {};
      \draw[cedge] ($(m)+(-1,1)$) -- (m) node[clabel,pos=0.3,above left] {\cmark};
      \draw[cedge] ($(m)+(1,1)$)  -- (m) node[clabel,pos=0.3,above right] {\cmark};
    \end{scope}
    \draw[-{Stealth[length=4mm]},very thick] (2,0) -- (3.4,0);
    \begin{scope}[xshift=5cm]
      \node[fg] (m2) {\cmark};
      \draw[cedge] ($(m2)+(-1,1)$) -- (m2) node[clabel,pos=0.3,above left] {\cmark};
      \draw[cedge] ($(m2)+(1,1)$)  -- (m2) node[clabel,pos=0.3,above right] {\cmark};
    \end{scope}
  \end{tikzpicture}
  \caption{Inferring convergence of a node using the merge rule}
\end{figure}

As shown on the left hand side of the above figure we have a merge node with an
unknown state, and exactly 2 predecessor edges which are known to be convergent.
We can infer that the node must be convergent as well.

\begin{figure}[h]
  \centering
  \begin{tikzpicture}[baseline]
    \begin{scope}
      \node[fg] (m) {\cmark};
      \draw[cedge] ($(m)+(-1,1)$) -- (m) node[clabel,pos=0.3,above left] {\cmark};
      \draw[uedge,very thick] ($(m)+(1,1)$) -- (m);
    \end{scope}
    \draw[-{Stealth[length=4mm]},very thick] (2,0) -- (3.4,0);
    \begin{scope}[xshift=5cm]
      \node[fg] (m2) {\cmark};
      \draw[cedge] ($(m2)+(-1,1)$) -- (m2) node[clabel,pos=0.3,above left] {\cmark};
      \draw[cedge] ($(m2)+(1,1)$)  -- (m2) node[clabel,pos=0.3,above right] {\cmark};
    \end{scope}
  \end{tikzpicture}
  \caption{Inferring convergence of an edge using the merge rule}
\end{figure}
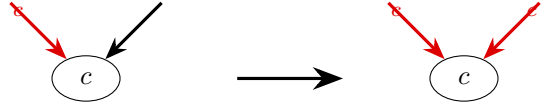

Again, in the above figure, on the left we have a merge node known to be
convergent and of its 2 predecessor edges only one is known to be convergent.
As shown on the right hand side of the figure we can assume the state of the
unknown edge to be convergent as well.

Again we note that the merge rule can be used to infer the state of a singleton
predecessor edge of a node as well. The following figure illustrates this idea.

\begin{figure}[h]
  \centering
  \begin{tikzpicture}[baseline]
    \begin{scope}
      \node[fg] (m) {};
      \draw[cedge] ($(m)+(0,1.2)$) -- (m) node[clabel,pos=0.35,right] {\cmark};
    \end{scope}
    \draw[-{Stealth[length=4mm]},very thick] (1.5,0) -- (2.9,0);
    \begin{scope}[xshift=4.4cm]
      \node[fg] (m2) {\cmark};
      \draw[cedge] ($(m2)+(0,1.2)$) -- (m2) node[clabel,pos=0.35,right] {\cmark};
    \end{scope}
  \end{tikzpicture}
  \caption{Inferring convergence of a singleton merge node using the merge rule}
\end{figure}

Similar to what we noted earlier, the merge rule propagates forward convergence
information from a set of known edges to its common successor merge node or
backwards from a merge node to one of its unknown predecessor edges.

\section{Convergence Analysis}

In general, convergence analysis is a bi-directional propagation problem. In
this section we describe an iterative work-list algorithm for propagating
convergence information in a flow graph and outline several improvements to
improve its precision.

\subsection{Basic Algorithm}

The key idea is to maintain a work list of nodes and edges with the property
that the state of the nodes and edges in the work list is solved trivially to
be true. As we remove elements from the work-list we determine if branch rule
or the merge rule can be used to determine new singleton unknown edges or nodes
and we add them to the work list.

The work list is initialized by the entry node, the exit node and the nodes
which are the barrier nodes of the flow graph. A well formed program must be
synchronized at the entry and the exit and we assume that all explicit
occurrences of the barrier statements are well synchronized or convergent.

Here is the sketch of the algorithm.

\begin{figure}[h]
\centering
\begin{minipage}{0.92\linewidth}
\begin{lstlisting}[style=algo]
// initialize
//
WorkList = NULL;
for each node in FlowGraph {
  state(node) = Unknown;
  switch kind(node) {
    case entry:
    case exit:
    case barrier:
      WorkList += node;
      break;
    default:
      break
  }
}
for each edge in FlowGraph {
  state(edge) = Unknown;
}

// solve
//
while (element = pop(WorkList) {
  state(element) = Convergent;
  switch (kind(element)) {
    case node:
      if (next = SingleBranchOrMergeRuleUnkown(element)) {
        WorkList += next;
      }
      break;
    case edge:
      if (next = SingleBranchOrMergeRuleUnknown(source(element))) {
        WorkList += next;
      }
      if (next = SingleBranchOrMergeRuleUnknown(target(element))) {
        WorkList += next;
      }
      break;
  }
}
\end{lstlisting}
\end{minipage}
\caption{Solver for Convergence Analysis}
\end{figure}

This algorithm runs in time linear in the number of edges and nodes of the flow
graph, in the worst case. The expected running time should be the number of
unknowns.

\subsection{Example}

In this section we give several examples on the results produced by the basic
algorithm.

\subsubsection{Acyclic Flow Graph}

Consider the following flow graph.

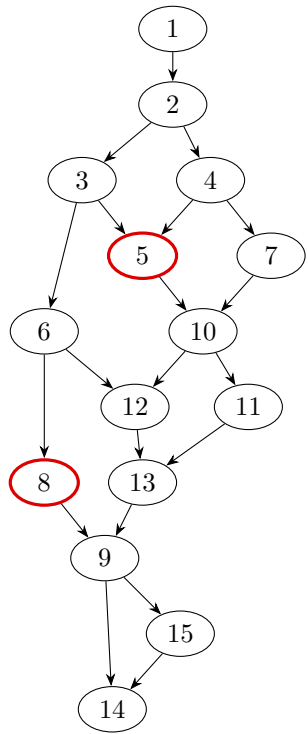
\begin{figure}[h]
  \centering
\begin{tikzpicture}[x=1cm,y=1cm]
  \node[fg]    (n1)  at ( 0.0,  0.0) {1};
  \node[fg]    (n2)  at ( 0.0, -1.0) {2};
  \node[fg]    (n3)  at (-1.2, -2.0) {3};
  \node[fg]    (n4)  at ( 0.5, -2.0) {4};
  \node[fgred] (n5)  at (-0.4, -3.0) {5};
  \node[fg]    (n7)  at ( 1.3, -3.0) {7};
  \node[fg]    (n6)  at (-1.7, -4.0) {6};
  \node[fg]    (n10) at ( 0.4, -4.0) {10};
  \node[fg]    (n12) at (-0.5, -5.0) {12};
  \node[fg]    (n11) at ( 1.0, -5.0) {11};
  \node[fgred] (n8)  at (-1.7, -6.0) {8};
  \node[fg]    (n13) at (-0.4, -6.0) {13};
  \node[fg]    (n9)  at (-0.9, -7.0) {9};
  \node[fg]    (n15) at ( 0.1, -8.0) {15};
  \node[fg]    (n14) at (-0.8, -9.0) {14};

  \draw[uedge] (n1)  -- (n2);
  \draw[uedge] (n2)  -- (n3);
  \draw[uedge] (n2)  -- (n4);
  \draw[uedge] (n3)  -- (n5);
  \draw[uedge] (n3)  -- (n6);
  \draw[uedge] (n4)  -- (n5);
  \draw[uedge] (n4)  -- (n7);
  \draw[uedge] (n5)  -- (n10);
  \draw[uedge] (n7)  -- (n10);
  \draw[uedge] (n6)  -- (n12);
  \draw[uedge] (n6)  -- (n8);
  \draw[uedge] (n10) -- (n12);
  \draw[uedge] (n10) -- (n11);
  \draw[uedge] (n12) -- (n13);
  \draw[uedge] (n11) -- (n13);
  \draw[uedge] (n8)  -- (n9);
  \draw[uedge] (n13) -- (n9);
  \draw[uedge] (n9)  -- (n15);
  \draw[uedge] (n9)  -- (n14);
  \draw[uedge] (n15) -- (n14);
\end{tikzpicture}
  \caption{An example flow graph}
  \label{fig:acyclic}
\end{figure}

The nodes 5 and 8 are the locations of the barrier statements, nodes 1 and 14
are the entry and exit nodes of the flow graph. After the basic algorithm
finishes the flow graph will be as follows with color red signifying places
that are convergent.

\begin{figure}[h]
  \centering
\begin{tikzpicture}[x=1cm,y=1cm]
  \node[fgred] (n1)  at ( 0.0,  0.0) {1};
  \node[fgred] (n2)  at ( 0.0, -1.0) {2};
  \node[fg]    (n3)  at (-1.2, -2.0) {3};
  \node[fg]    (n4)  at ( 0.5, -2.0) {4};
  \node[fgred] (n5)  at (-0.4, -3.0) {5};
  \node[fg]    (n7)  at ( 1.3, -3.0) {7};
  \node[fg]    (n6)  at (-1.7, -4.0) {6};
  \node[fg]    (n10) at ( 0.4, -4.0) {10};
  \node[fg]    (n12) at (-0.5, -5.0) {12};
  \node[fg]    (n11) at ( 1.0, -5.0) {11};
  \node[fgred] (n8)  at (-1.7, -6.0) {8};
  \node[fg]    (n13) at (-0.4, -6.0) {13};
  \node[fg]    (n9)  at (-0.9, -7.0) {9};
  \node[fg]    (n15) at ( 0.1, -8.0) {15};
  \node[fgred] (n14) at (-0.8, -9.0) {14};

  \draw[cedge] (n1)  -- (n2);
  \draw[uedge] (n2)  -- (n3);
  \draw[uedge] (n2)  -- (n4);
  \draw[uedge] (n3)  -- (n5);
  \draw[uedge] (n3)  -- (n6);
  \draw[uedge] (n4)  -- (n5);
  \draw[uedge] (n4)  -- (n7);
  \draw[cedge] (n5)  -- (n10);
  \draw[uedge] (n7)  -- (n10);
  \draw[uedge] (n6)  -- (n12);
  \draw[cedge] (n6)  -- (n8);
  \draw[uedge] (n10) -- (n12);
  \draw[uedge] (n10) -- (n11);
  \draw[uedge] (n12) -- (n13);
  \draw[uedge] (n11) -- (n13);
  \draw[cedge] (n8)  -- (n9);
  \draw[uedge] (n13) -- (n9);
  \draw[uedge] (n9)  -- (n15);
  \draw[uedge] (n9)  -- (n14);
  \draw[uedge] (n15) -- (n14);
\end{tikzpicture}
  \caption{An example flow graph with convergence information}
\end{figure}
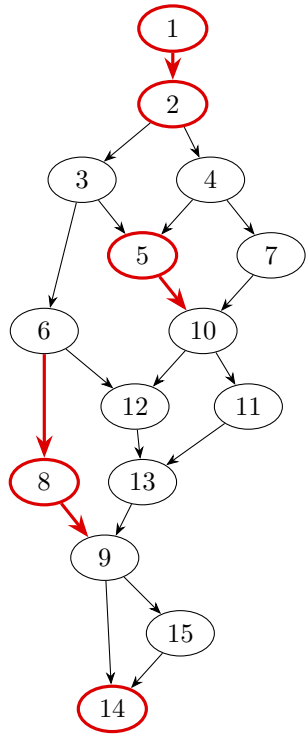

\subsubsection{Flowgraph with Cycles}

In this section we show an example with loops.

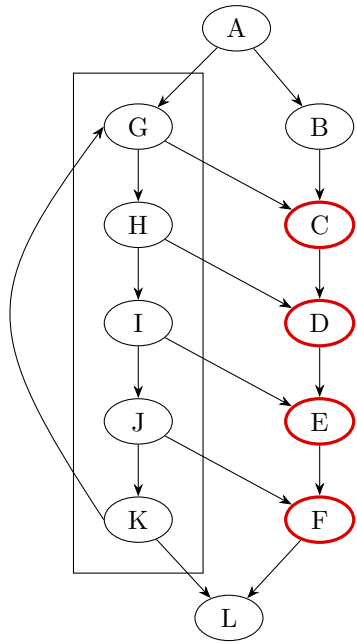
\begin{figure}[h]
  \centering
\begin{tikzpicture}[x=1cm,y=1cm]
  \node[fg]    (A) at ( 0.2,  0.0) {A};
  \node[fg]    (G) at (-1.1, -1.3) {G};
  \node[fg]    (B) at ( 1.3, -1.3) {B};
  \node[fg]    (H) at (-1.1, -2.6) {H};
  \node[fgred] (C) at ( 1.3, -2.6) {C};
  \node[fg]    (I) at (-1.1, -3.9) {I};
  \node[fgred] (D) at ( 1.3, -3.9) {D};
  \node[fg]    (J) at (-1.1, -5.2) {J};
  \node[fgred] (E) at ( 1.3, -5.2) {E};
  \node[fg]    (K) at (-1.1, -6.5) {K};
  \node[fgred] (F) at ( 1.3, -6.5) {F};
  \node[fg]    (L) at ( 0.1, -7.8) {L};

  \begin{scope}[on background layer]
    \node[draw, fit=(G)(H)(I)(J)(K), inner sep=4mm] {};
  \end{scope}

  \draw[uedge] (A) -- (G);
  \draw[uedge] (A) -- (B);
  \draw[uedge] (B) -- (C);
  \draw[uedge] (G) -- (H);
  \draw[uedge] (G) -- (C);
  \draw[uedge] (H) -- (I);
  \draw[uedge] (H) -- (D);
  \draw[uedge] (I) -- (J);
  \draw[uedge] (I) -- (E);
  \draw[uedge] (J) -- (K);
  \draw[uedge] (J) -- (F);
  \draw[uedge] (K) -- (L);
  \draw[uedge] (C) -- (D);
  \draw[uedge] (D) -- (E);
  \draw[uedge] (E) -- (F);
  \draw[uedge] (F) -- (L);
  \draw[uedge] (K.west) .. controls (-3.2,-4.0) and (-3.2,-3.5) .. (G.west);
\end{tikzpicture}
  \caption{An example flow graph with loop}
\end{figure}

This example contains a loop with several exits. The nodes C, D, E, and F are
barrier statements, and nodes A and L are entry and exit nodes, respectively.
The analysis result from our algorithm is shown below.

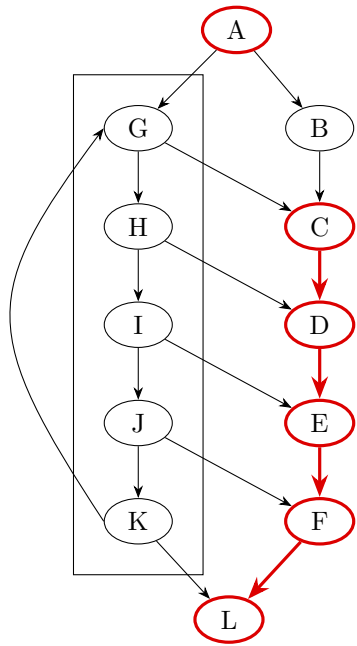
\begin{figure}[h]
  \centering
\begin{tikzpicture}[x=1cm,y=1cm]
  \node[fgred] (A) at ( 0.2,  0.0) {A};
  \node[fg]    (G) at (-1.1, -1.3) {G};
  \node[fg]    (B) at ( 1.3, -1.3) {B};
  \node[fg]    (H) at (-1.1, -2.6) {H};
  \node[fgred] (C) at ( 1.3, -2.6) {C};
  \node[fg]    (I) at (-1.1, -3.9) {I};
  \node[fgred] (D) at ( 1.3, -3.9) {D};
  \node[fg]    (J) at (-1.1, -5.2) {J};
  \node[fgred] (E) at ( 1.3, -5.2) {E};
  \node[fg]    (K) at (-1.1, -6.5) {K};
  \node[fgred] (F) at ( 1.3, -6.5) {F};
  \node[fgred] (L) at ( 0.1, -7.8) {L};

  \begin{scope}[on background layer]
    \node[draw, fit=(G)(H)(I)(J)(K), inner sep=4mm] {};
  \end{scope}

  \draw[uedge] (A) -- (G);
  \draw[uedge] (A) -- (B);
  \draw[uedge] (B) -- (C);
  \draw[uedge] (G) -- (H);
  \draw[uedge] (G) -- (C);
  \draw[uedge] (H) -- (I);
  \draw[uedge] (H) -- (D);
  \draw[uedge] (I) -- (J);
  \draw[uedge] (I) -- (E);
  \draw[uedge] (J) -- (K);
  \draw[uedge] (J) -- (F);
  \draw[uedge] (K) -- (L);
  \draw[cedge] (C) -- (D);
  \draw[cedge] (D) -- (E);
  \draw[cedge] (E) -- (F);
  \draw[cedge] (F) -- (L);
  \draw[uedge] (K.west) .. controls (-3.2,-4.0) and (-3.2,-3.5) .. (G.west);
\end{tikzpicture}
  \caption{Result of convergence analysis on loops}
\end{figure}

It is interesting to note that the loop back edge and the first loop exit
$G \rightarrow C$ cannot be proved convergent. However, if the loop back edge is
removed than the analysis will be able to show more convergent edges and nodes
as follows.

\begin{figure}[h]
  \centering
\begin{tikzpicture}[x=1cm,y=1cm]
  \node[fgred] (A) at ( 0.2,  0.0) {A};
  \node[fgred] (G) at (-1.1, -1.3) {G};
  \node[fg]    (B) at ( 1.3, -1.3) {B};
  \node[fgred] (H) at (-1.1, -2.6) {H};
  \node[fgred] (C) at ( 1.3, -2.6) {C};
  \node[fgred] (I) at (-1.1, -3.9) {I};
  \node[fgred] (D) at ( 1.3, -3.9) {D};
  \node[fgred] (J) at (-1.1, -5.2) {J};
  \node[fgred] (E) at ( 1.3, -5.2) {E};
  \node[fgred] (K) at (-1.1, -6.5) {K};
  \node[fgred] (F) at ( 1.3, -6.5) {F};
  \node[fgred] (L) at ( 0.1, -7.8) {L};

  \begin{scope}[on background layer]
    \node[draw, fit=(G)(H)(I)(J)(K), inner sep=4mm] {};
  \end{scope}

  \draw[cedge] (A) -- (G);
  \draw[uedge] (A) -- (B);
  \draw[uedge] (B) -- (C);
  \draw[cedge] (G) -- (H);
  \draw[cedge] (G) -- (C);
  \draw[cedge] (H) -- (I);
  \draw[cedge] (H) -- (D);
  \draw[cedge] (I) -- (J);
  \draw[cedge] (I) -- (E);
  \draw[cedge] (J) -- (K);
  \draw[cedge] (J) -- (F);
  \draw[cedge] (K) -- (L);
  \draw[cedge] (C) -- (D);
  \draw[cedge] (D) -- (E);
  \draw[cedge] (E) -- (F);
  \draw[cedge] (F) -- (L);
\end{tikzpicture}
  \caption{Result of convergence analysis by removing loop backedge}
\end{figure}
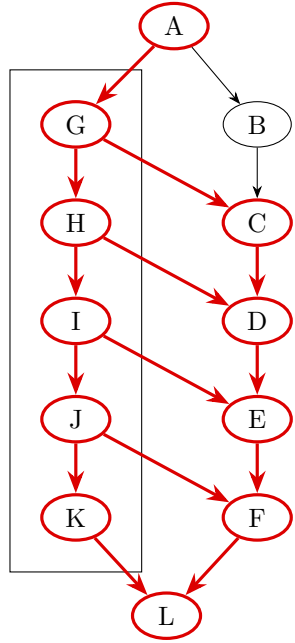

\subsection{Using Region Information}

In the previous \hyperref[fig:acyclic]{example}, we notice that there are
several nodes that are not discovered to be convergent, even though they are.
For example, nodes 2 and node 9 are entry and exit, respectively, of a single
entry single exit region and therefore the convergence information at the two
nodes must be the same.

In the following we define a pair of distinct nodes, say $(x, y)$, as forming a
region entry-exit pair if
\begin{enumerate}[leftmargin=2.2em]
  \item $x$ strictly dominates $y$ and $y$ strictly post-dominates $x$, and
  \item $x$ and $y$ belong to the same interval.
\end{enumerate}

These 2 conditions guarantee that if a region entry exit pair satisfies the 2
given conditions then their convergence states are identical.

This insight allows us to make an important improvement to the solver. The
basic idea is to add the region entry and region exit nodes, to the work list.
The improved algorithm is shown below.

\begin{figure}[h]
\centering
\begin{minipage}{0.92\linewidth}
\begin{lstlisting}[style=algo]
// solve
//
while (element = pop(WorkList) {
  state(element) = Convergent;
  switch (kind(element)) {
    case node:
      if (next = SingleBranchOrMergeRuleUnkown(element)) {
        WorkList += next;
      }
      if (next = RegionEntryOrExit(element)) {
        WorkList += next;
      }
      break;
    case edge:
      if (next = SingleBranchOrMergeRuleUnknown(source(element))) {
        WorkList += next;
      }
      if (next = SingleBranchOrMergeRuleUnknown(target(element))) {
        WorkList += next;
      }
      break;
  }
}
\end{lstlisting}
\end{minipage}
\caption{Region based Solver for Convergence Analysis}
\end{figure}

With this improvement the algorithm is able to discover the following
information.

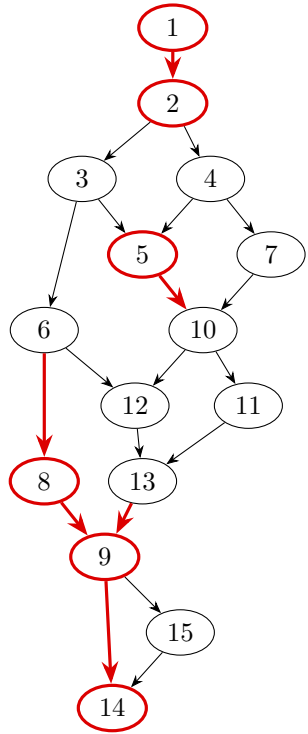
\begin{figure}[h]
  \centering
\begin{tikzpicture}[x=1cm,y=1cm]
  \node[fgred] (n1)  at ( 0.0,  0.0) {1};
  \node[fgred] (n2)  at ( 0.0, -1.0) {2};
  \node[fg]    (n3)  at (-1.2, -2.0) {3};
  \node[fg]    (n4)  at ( 0.5, -2.0) {4};
  \node[fgred] (n5)  at (-0.4, -3.0) {5};
  \node[fg]    (n7)  at ( 1.3, -3.0) {7};
  \node[fg]    (n6)  at (-1.7, -4.0) {6};
  \node[fg]    (n10) at ( 0.4, -4.0) {10};
  \node[fg]    (n12) at (-0.5, -5.0) {12};
  \node[fg]    (n11) at ( 1.0, -5.0) {11};
  \node[fgred] (n8)  at (-1.7, -6.0) {8};
  \node[fg]    (n13) at (-0.4, -6.0) {13};
  \node[fgred] (n9)  at (-0.9, -7.0) {9};
  \node[fg]    (n15) at ( 0.1, -8.0) {15};
  \node[fgred] (n14) at (-0.8, -9.0) {14};

  \draw[cedge] (n1)  -- (n2);
  \draw[uedge] (n2)  -- (n3);
  \draw[uedge] (n2)  -- (n4);
  \draw[uedge] (n3)  -- (n5);
  \draw[uedge] (n3)  -- (n6);
  \draw[uedge] (n4)  -- (n5);
  \draw[uedge] (n4)  -- (n7);
  \draw[cedge] (n5)  -- (n10);
  \draw[uedge] (n7)  -- (n10);
  \draw[uedge] (n6)  -- (n12);
  \draw[cedge] (n6)  -- (n8);
  \draw[uedge] (n10) -- (n12);
  \draw[uedge] (n10) -- (n11);
  \draw[uedge] (n12) -- (n13);
  \draw[uedge] (n11) -- (n13);
  \draw[cedge] (n8)  -- (n9);
  \draw[cedge] (n13) -- (n9);
  \draw[uedge] (n9)  -- (n15);
  \draw[cedge] (n9)  -- (n14);
  \draw[uedge] (n15) -- (n14);
\end{tikzpicture}
  \caption{Convergence information from region inference}
\end{figure}

\subsection{Using Path Information}

Consider the following flow graph.

\begin{figure}[h]
  \centering
\begin{tikzpicture}[x=1cm,y=1cm]
  \node[fg]    (X)    at ( 0.0,  0.0) {X};
  \node[fg]    (Cond) at ( 0.0, -1.2) {Cond};
  \node[fgred] (A)    at (-1.3, -2.4) {A};
  \node[fg]    (B)    at ( 0.5, -2.4) {B};
  \node[fg]    (F)    at (-1.3, -3.6) {F};
  \node[fg]    (C)    at ( 0.2, -3.6) {C};
  \node[fgred] (E)    at (-0.5, -4.8) {E};
  \node[fg]    (D)    at ( 1.0, -4.8) {D};
  \node[fg]    (Y)    at (-0.2, -6.0) {Y};

  \draw[uedge] (X)    -- (Cond);
  \draw[uedge] (Cond) -- (A);
  \draw[uedge] (Cond) -- (B);
  \draw[uedge] (A)    -- (F);
  \draw[uedge] (B)    -- (C);
  \draw[uedge] (B)    -- (D);
  \draw[uedge] (C)    -- (E);
  \draw[uedge] (C)    -- (D);
  \draw[uedge] (F)    -- (E);
  \draw[uedge] (F)    -- (D);
  \draw[uedge] (E)    -- (Y);
  \draw[uedge] (D)    -- (Y);
\end{tikzpicture}
  \caption*{}
\end{figure}

In this figure assume that nodes X and Y are the entry and exit nodes,
respectively. Nodes A and E are the barrier nodes. Our algorithm will discover
the following:

\begin{figure}[h]
  \centering
\begin{tikzpicture}[x=1cm,y=1cm]
  \node[fgred] (X)    at ( 0.0,  0.0) {X};
  \node[fgred] (Cond) at ( 0.0, -1.2) {Cond};
  \node[fgred] (A)    at (-1.3, -2.4) {A};
  \node[fgred] (B)    at ( 0.5, -2.4) {B};
  \node[fgred] (F)    at (-1.3, -3.6) {F};
  \node[fg]    (C)    at ( 0.2, -3.6) {C};
  \node[fgred] (E)    at (-0.5, -4.8) {E};
  \node[fg]    (D)    at ( 1.0, -4.8) {D};
  \node[fgred] (Y)    at (-0.2, -6.0) {Y};

  \draw[cedge] (X)    -- (Cond);
  \draw[cedge] (Cond) -- (A);
  \draw[cedge] (Cond) -- (B);
  \draw[cedge] (A)    -- (F);
  \draw[uedge] (B)    -- (C);
  \draw[uedge] (B)    -- (D);
  \draw[uedge] (C)    -- (E);
  \draw[uedge] (C)    -- (D);
  \draw[uedge] (F)    -- (E);   
  \draw[uedge] (F)    -- (D);
  \draw[cedge] (E)    -- (Y);
  \draw[uedge] (D)    -- (Y);
\end{tikzpicture}
  \caption*{}
\end{figure}

It should be obvious that the edge F to E is convergent but it is not
discovered, so the question is how can we deduce that?

Intuitively the rule for discovering this may be as follows. If there is an
edge between 2 convergent nodes (say $X \rightarrow Y$) and all paths from X to
Y go through that edge then that edge can be assumed to be convergent. Is this
correct? How can we formalize this?

\subsection{Using Variance Information}

So far we have not used any data flow information in the algorithm. All
information is derived from the flow graph structure and assumption that the
program is well synchronized. Can we incorporate variance information in this
analysis?

One way we could incorporate this into our analysis might be as follows. If a
branch node is convergent and its branch condition is thread invariant then all
successor edges of the branch node can be assumed as convergent.

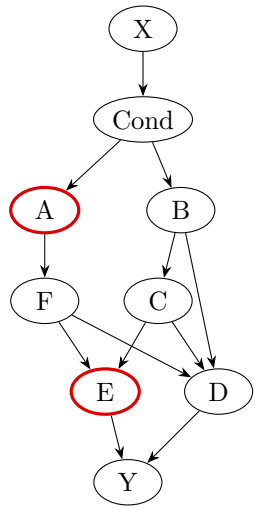
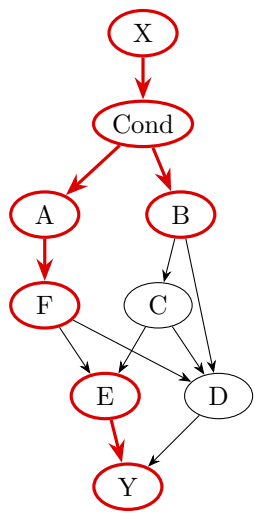
\end{document}